\documentclass[letterpaper,11pt]{article}
\pdfoutput=1

\usepackage{cleveref}
\usepackage{jheppub}
\usepackage{booktabs} 
\usepackage{graphicx} 
\usepackage{siunitx}
\usepackage{hyperref}
\usepackage{epsfig}
\usepackage[normalem]{ulem}
\usepackage{bm}
\usepackage{bbm}
\usepackage{slashed}
\usepackage{xspace}
\usepackage{subfigure}
\usepackage{multirow}

\usepackage{amsmath,latexsym}
\usepackage{pstricks}
\usepackage{color} 
 
\usepackage{marginnote}

\newcommand{\nn}{\nonumber}

\newcommand{\beq}{\begin{equation}}
\newcommand{\eeq}{\end{equation}}
\newcommand{\bea}{\begin{eqnarray}}
\newcommand{\eea}{\end{eqnarray}}

\begin{document}


\title{Virtual Hawking Radiation}

\author[a]{Walter D. Goldberger,}
\author[b]{Ira Z. Rothstein}
\affiliation[a]{Department of Physics, Yale University, New Haven, CT 06511}
\affiliation[b]{Department of Physics, Carnegie Mellon University, Pittsburgh, PA 15213}

\vspace{0.3cm}


\abstract{

We consider the effects of off-shell Hawking radiation on scattering processes involving black holes coupled to quantum fields.   The focus here is to the case of gravitational scattering of a scalar field mediated by the exchange of virtual Hawking gravitons from a four-dimensional Schwarzschild black hole.   Our result is obtained in the context of a worldline effective field theory for the black hole, and is valid in the semi-classical limit where the Schwarzschild radius $r_s$ is larger than the Planck length $1/m_{Pl}$.    In addition, we assume that four-momentum exchange $q$ is smaller than $r_s^{-1}$ and that the incoming particle has energy larger then the black hole's Hawking temperature.    The inelastic cross section we obtain is a new, leading order quantum gravity effect, arising at the same order in $q^2/m_{Pl}^2$ as the well understood one-loop graviton vacuum polarization corrections to gravitational scattering between massive particles.

}
\maketitle



\section{Introduction}
While quantum gravity at the Planck scale still remains a mystery, it is commonly believed that the low-energy gravitational $S$-matrix is by now completely understood.    In particular, the UV divergences that arise in the calculation of scattering amplitudes containing either on-shell or virtual gravitons can be treated systematically by the standard tools of effective field theory (EFT) \cite{donoghue}.    For instance, the graviton-graviton elastic scattering amplitude in pure gravity is finite at one-loop~\cite{tv}, while at two loops~\cite{Goroff:1985th} it exhibits logarithmic divergences which can be absorbed into the Einstein-Hilbert action supplemented by higher-derivative operators, schematically\footnote{We define $m_{Pl}^2 =1/32\pi G_N$.    In Eq.~(\ref{eq:gs}), curvature squared terms can be either removed by field redefinitions of the graviton or traded for topological terms which have no effect on perturbative observables.}
\beq
\label{eq:gs}
S= - 2 m_{Pl}^2 \int d^4 x \sqrt{g} \left[R + {1\over m_{Pl}^4} R_{\mu\nu\alpha\beta}^3+\cdots\right],
\eeq
with higher order terms suppressed by more powers of $E^2/m_{Pl}^2\ll1$ at low energies. 

Although this methodology yields well-defined long distance predictions of quantum gravity for processes involving elementary particles coupled to gravitons, the computation of the quantum gravity $S$-matrix with black hole asymptotic states\footnote{The meaning of ``asymptotic state'' for the case of black holes which decay via the emission of Hawking radiation will be further discussed in sec.~\ref{sec:EFTsetup}} has yet to be accomplished, even at energies below the Planck scale.    In this case, there are new non-perturbative quantum effects associated with Hawking radiation~\cite{Hawking:1974sw}, which play a crucial role.    The emission of on-shell Hawking radiation from fixed black hole backgrounds has been thoroughly studied~\cite{page2}.  However, the effects of \emph{virtual} Hawking modes represent another source of quantum gravity corrections whose analysis is still unchartered territory that should be amenable to a field theoretic treatment.

In a recent paper~\cite{GnR3}, we have introduced an effective field theory framework designed to calculate quantum corrections to processes involving black holes interacting through the exchange of long wavelength fields.   It builds on methods described in~\cite{GnR2,GnR4} to treat the effects of classical absorption by the horizon of a black hole within a world-line effective theory  \cite{GnR1} of the gravitational dynamics of compact objects.   The main idea of~\cite{GnR3,GnR2,GnR4} is that emission and absorption of quanta by the horizon are due to horizon localized degrees of freedom which couple to external fields.   In the limit where the black hole radius $r_s= 2 G_N M$ is small, these localized modes are described by a quantum mechanical ($0+1$) theory whose correlation functions can be extracted model-independently, by matching to on-shell emission and absorption processes in the full semi-classical black hole spacetime.   The same correlators can then be used to predict observables where the black hole horizon exchanges off-shell modes with other objects, for instance the classical dissipation of energy in the binary dynamics of comparable mass black holes~\cite{GnR2,GnR4}.

A somewhat counterinuitve property of the black hole worldline correlators obtained in~\cite{GnR3} is that, for black holes in the Unruh state~\cite{Unruh:1976db} (i.e. black holes formed from the gravitational collapse of matter), Hawking emission is not suppressed by powers of $\hbar$.  Instead the Hawking response is enhanced at low frequency relative to classical absorption by the horizon.   This is tied to the well-known fact the the distribution of emitted Hawking quanta from a semi-classical black hole is independent of $\hbar$, as well as to detailed balance arguments for black holes in thermal equilibrium~\cite{Hartle:1976tp} with a radiation bath at the Hawking temperature $T_H =\hbar/4\pi r_s$. While the Wightman functions calculated in~\cite{GnR3} are themselves not suppressed by powers of $\hbar$ or, equivalently, $1/m_{Pl}$, the causal (retarded) Green's functions in the Unruh state were shown to be insensitive to Hawking radiation, at least up to corrections from bulk interactions of the fields propagating around the black hole.  As a consequence, there are no observable corrections to classical processes (e.g. binary dynamics) from Hawking modes, as classical intuition would suggest.

On the other hand, there are processes, such as quantum mechanical scattering of matter fields incident on the black hole, which depend on worldline correlators other than the causal two-point function.    In this case, the effects of (off-shell) Hawking radiation do not cancel.  It is then interesting to ask how their magnitude compares to the more familiar loop corrections based on a perturbative treatment of Eq.~(\ref{eq:gs}).   To address this, in this paper we consider an inelastic scattering process where a quantum field (for simplicity, a complex scalar $\phi$) minimally coupled to gravity scatters off a black hole via the exchange of an off-shell Hawking graviton mode.    We obtain a well-defined (calculable) prediction for the inelastic scattering cross section, which is of the \emph{same} order in $1/m^2_{Pl}$ as the canonical one-loop quantum gravity corrections to the elastic scattering cross section  corrections arising from interference with single graviton exchange.

In sec.~\ref{sec:EFTsetup}, we summarize the EFT setup, including the relevant hierarchy of scales in which our description holds as well as the systematics of the power counting.   Details of the matching calculation needed to extract the relevant worldline correlators can be found in the appendix~\ref{app}.  In sec.~\ref{sec:scattering}, we compute the leading order inelastic process induced by Hawking graviton exchange and compare to the elastic scattering process.   Our main result is given in Eq.~(\ref{eq:result}).    Finally, in sec.~\ref{sec:conc} we summarize and outline directions for future work.

\section{The EFT formalism and power counting}
\label{sec:EFTsetup}

We are interested in scattering processes where matter fields scatter gravitationally off a quantum mechanical black hole.   To be definite, we consider the case of a complex scalar field $\phi$ coupled minimally to gravity,
\beq
\label{eq:S}
S = \int d^4x \sqrt{g} \left(g^{\mu \nu} \partial_\mu \phi^\dagger \partial_\nu \phi -m^2 \phi^\dagger \phi\right).
\eeq
This action, along with the Einstein-Hilbert term $S_{EH}=-2 m_{Pl}^2\int d^4 x \sqrt{g} R+\cdots$, is sufficient to study the effects of quantum gravity as long as we are interested in processes where all energy and momentum scales, and therefore the curvature, are small compared to the Planck scale.   

Of particular interest here is the case where the scalar $\phi$ and the graviton field $h_{\mu\nu}$ propagate in the background of a black hole solution to Einstein's equations.   We take the case of Schwarzschild black holes for simplicity, and assume that the curvature at the horizon is small in Planck units.   Then the interactions of scalar and graviton can be analyzed in a derivative expansion of the action about the Schwarzschild background that is both systematic and  tractable.    

In order to sidestep the technical difficulties of quantizing the graviton in the full Schwarzschild background, including the effects of Hawking radiation from the black hole horizon, we will use the effective field theory methods developed in \cite{GnR1,GnR2,GnR3,GnR4}.   In this EFT one begins by first considering the black hole in the point particle approximation.  In so doing we have integrated out all of the internal dynamics, with finite size effects systematically accounted for by including all higher dimensional operators (composed of the curvature, as well as other fields) that are consistent with symmetries of the underlying UV theory.    This will not suffice, however, to describe either Hawking radiation or absorption, which imply the existence of gapless degrees of freedom associated with the dynamics of the horizon.

To account for these gapless modes in a model independent way, we introduce a quantum mechanical Hilbert space of states localized on the black hole worldline coordinate $x^\mu(\tau)$.  In this description the semi-classical black hole with mass $M\gg m_{Pl}$ corresponds to a highly excited state $|M\rangle$ where the mass is hierarchically larger than the gap between states, of order $1/r_s$.   In the absence of couplings to, e.g., external gravitational or electromagnetic interactions, the state $|M\rangle$ is an eigenstate of the black hole Hamiltonian $H_o$.    The external fields couple to composite worldline operators made out of the black hole internal degrees of freedom.  Absent a specific model, we classify these operators by their quantum numbers under $SO(3)$ isometries of the Schwarzschild geometry, and couple them to external fields in all ways consistent with symmetry.   For instance, at leading order in the multipole expansion, the tidal gravitational response is accounted for by including  $\ell=2$ (quadrupole) operators $Q^E_{ab}(\tau)$,  $Q^B_{ab}(\tau)$ of electric and magnetic parity, whose gravitational interactions are encoded in the action
\beq
\label{eq:EFTS}
S_{int} = -\int d\tau Q^E_{ab}(\tau) E^{ab}(x(\tau))-\int d\tau(\tau) Q^B_{ab}(\tau) B^{ab}(x(\tau)).
\eeq 
Here, the indices $a,b=1,2,3$ refer to a spatial frame $e^a_\mu(\tau)$ that describes the orientation of the black hole relative to the ambient space.   By definition this frame obeys the constraints $v^\mu e^a_\mu =0$, and   
\begin{eqnarray}
\nonumber
g^{\mu\nu} e^a_\mu e^a_\nu &=& - \delta^{ab}, \\
 \delta_{ab} e^a_\mu e^b_\nu &=& g_{\mu\nu} - v_\mu v_\nu,
\end{eqnarray}
with $v^\mu=dx^\mu/d\tau$ the four-velocity of the black hole.     The projected curvature tensors are $E^{ab} = e^a_\mu e^b_\nu E^{\mu\nu}$, and $B^{ab} = e^a_\mu e^b_\nu B^{\mu\nu}$, where the electric and magnetic components of the curvature tensor\footnote{In practice, the Ricci curvature parts of $R_{\mu\nu\alpha\beta}$ can be removed by field redefinitions of the graviton, so do not have any physical effects} are 
\bea
\nn
E_{\mu\nu} &=& R_{\mu\alpha\nu\beta} v^\alpha v^\beta,\\
B_{\mu\nu} &=&  {\tilde R}_{\mu\alpha\nu\beta}  v^\alpha v^\beta = {1\over 2} \epsilon_{\mu\alpha\rho\sigma} R^{\rho\sigma}{}_{\beta\nu}  v^\alpha v^\beta. 
\eea

The validity of the effective worldline description is limited to the regime where the black hole interacts with probes whose typical frequency (or wavenumber) $\omega$ lies in the range
\beq
\label{eq:limits}
\tau^{-1}_{BH} \ll \omega \ll 1/r_s, 
\eeq
where the upper bound arises as a consequence of the point particle approximation ($r_s=2 G_N M$ is the Schwarzschild radius), and the lower bound ensures that we are looking at time scales short compared to the Page time $\tau_{BH}\sim M^3/m_{Pl}^2$, so that we can ignore the backreaction due to the evaporation process.   We also take the black hole to be semi-classical, with mass $M\gg m_{Pl}$. 

As explained in~\cite{GnR2,GnR3}, physical processes involving the black hole coupled to other fields are described in this EFT in terms of the Wightman functions of worldline operators such as $Q^E_{ab}(\tau)$,  $Q^B_{ab}(\tau)$ which can be obtained by a matching calculation to the full theory of fields propagating in the black hole spacetime.   For a non-rotating black hole, the two-point Wightman functions in the frame where the black hole is at rest then take the form,
\beq
\label{eq:wight}
\langle M|Q^{E,B}_{ab}(t)  Q^{E,B}_{cd}(0)|M \rangle = \langle a,b|c,d\rangle \int_{-\infty}^{\infty} {d\omega\over 2\pi} e^{-i \omega t} A^{E,B}_+(\omega),
\eeq
where $\langle a,b|c,d\rangle= {1\over 2} \left[\delta_{ac} \delta_{bd} + \delta_{ad}\delta_{bc} - {2\over 3} \delta_{ab}\delta_{cd}\right]$ is the identity operator on the space of $\ell=2$ tensors.    

The Wightman function $A^{E,B}_+(\omega)$ are obtained by a matching calculation described in~\cite{GnR2}, which compares the EFT to multi-particle scattering and production probabilites~\cite{bek,wald} $p(n\rightarrow m)$ for the black hole in the Unruh state~\cite{Unruh:1976db} (corresponding to a non-eternal black hole, formed by realistic gravitational collapse).   Adapting the methods of~\cite{GnR2} to the case of gravitons, we find in the appendix that to leading order in $r_s\omega \ll 1$, 
\beq
\label{eq:aeab}
A_+^E(\omega) = A_+^B(\omega)\approx {r_s^5\over  360\pi G_N},
\eeq
In particular, the presence of non-vanishing response at $\omega<0$ accounts for emission of Hawking gravitons near the horizon, while the $\omega>0$ branch represents absorption.

In this paper, we will use the EFT to analyze the inelastic scattering of matter fields, represented here by the complex scalar $\phi$ of Eq.~(\ref{eq:S}) incident on a black hole with mass $M\gg m$.  Since the case where the scalar field has negligible mass compared to the Hawking temperature, $T_H=(4\pi r_s)^{-1}$,  corresponding to absorption and re-emission of on-shell scalars, is well understood \cite{bek}, we focus instead on the limit $m\gg T_H$.    In this regime, the dominant inelastic process is through the exchange of off-shell Hawking gravitons between the scalar and the black hole.   Alternatively, one could also study the limit where the incoming scalar has energy $E_\phi\gg T_H$, where again the scattering process is dominated by graviton exchange.   However, in order to remain within the regime of validity of the EFT, we take the typical momentum transfer $q$ (or equivalently, the impact parameter $b\sim 1/q$) to lie in the region defined by Eq.~(\ref{eq:limits}).   Thus to ensure the validity of the EFT, we assume the following hierarchy of kinematic scales 
\beq
M\gg E_\phi \gg T_H\gg q\sim b^{-1}\gg \tau_{BH}^{-1}.
\eeq

\section{Scattering by off-shell Hawking radiation}
\label{sec:scattering}

\begin{figure}
    \centering
    \includegraphics[scale=0.2]{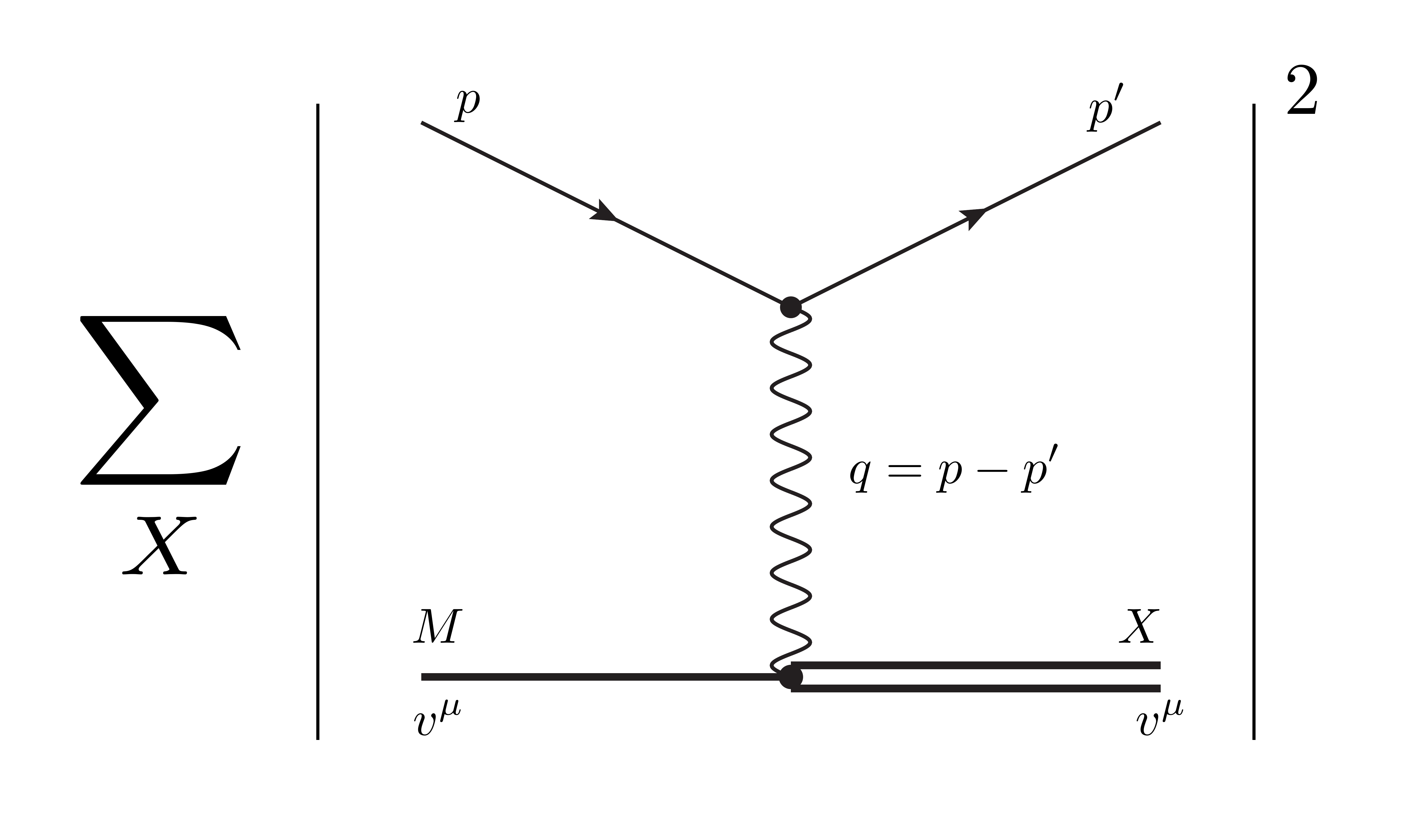}
\caption{Leading order gravitational inelastic scattering of a scalar field incident on a semi-classical Schwarzschild black hole.   }
\label{fig:amp}
\end{figure}

We now compute the inelastic process $\phi(p)+\mbox{BH}_M\rightarrow \phi(p') + X$ where a scalar field scatters off a heavy Schwarzschild black hole.    Due to the presence of the horizon, the scalar can tidally exchange energy and momentum with the black hole.    In the Unruh state, the exchanged energy can be of either sign, due to the possibility that $\phi$ absorbs a virtual Hawking mode emitted by the black hole.

In the EFT, the inclusive probability is given in Fig.~\ref{fig:amp}, where we sum over the unobserved internal states $X$ of the final black hole.   The interaction between the black hole and $\phi$ is mediated by graviton exchange.   Linearizing about flat space, $g_{\mu\nu}=\eta_{\mu\nu}+h_{\mu\nu}/m_{Pl}$, the relevant term in Eq.~(\ref{eq:S}) is
\beq
{\cal L}_{int} =-{ h^{\mu\nu}\over m_{Pl}} \left[\partial_\mu\phi^\dagger\partial_\nu \phi -{1\over 2} \eta_{\mu\nu} \left(\left|\partial\phi\right|^2-m^2\phi^\dagger \phi\right)\right].
\eeq
Our calculation is performed in Feynman gauge, where the propagator of the exchanged graviton  is ${i\over q^2} P_{\mu\nu;\alpha\beta}$, with
 \beq
 P_{\mu\nu;\alpha\beta}= {1\over 2}\left[\eta_{\mu\alpha}\eta_{\nu\beta}+\eta_{\mu\beta}\eta_{\nu\alpha}-\eta_{\mu\nu}\eta_{\alpha\beta}\right].
 \eeq
In the rest frame of the black hole $v^\mu=(1,0)$, the amplitude to leading order in the EFT of Eq.~(\ref{eq:EFTS}) takes the form ($q=p-p'$ is the momentum transfer)
 \beq
 {i\cal A}_X = -{1\over 2 m_{Pl}^2} \cdot {i\over q^2} \int d\tau e^{-iq\cdot v\tau}\langle X|Q^E_{ab}(\tau)|M\rangle {\cal A}^E_{ab} + \mbox{mag.}
\eeq
where the tensors ${\cal A}^E_{ab}=e^\mu{}_a e^\nu{}_b {\cal A}^E_{\mu\nu}$ are given by\footnote{$q^\mu_\perp = q^\mu - (v\cdot q) v^\mu$, $\eta^{\mu\nu}_\perp=\eta^{\mu\nu}-v^\mu v^\nu$.} 
\bea
\label{eq:AE}
{\cal A}^E_{\mu\nu} &=& \left[(v\cdot p) q - (v\cdot q) p\right]^\mu \left[(v\cdot p) q - (v\cdot q) p\right]^\nu -{1\over 2} m^2\left[q^\perp_\mu q^\perp_\nu + (v\cdot q)^2 \eta_{\mu\nu}^\perp\right] ,\\
\label{eq:AB}
 {\cal A}^B_{\mu\nu} &=&  \epsilon_{\mu\alpha\rho\sigma} v^\alpha p^\rho q^\sigma \left[(v\cdot q) p - (v\cdot p) q\right]^\nu +{1\over 2} m^2 (v\cdot q) \epsilon^{\mu\nu\rho\sigma} v_\rho q_\sigma.
 \eea
 In order to perform the tensor contractions, we have used the package~\cite{Mertig:1990an}.   
 
 Summing over the final states $X$, and assuming unitarity of the black hole quantum mechanics, $\sum_X |X\rangle \langle X| = 1$, the inclusive squared amplitude breaks up into electric and magnetic contributions
  \beq
 \sum_X |{\cal A}_X|^2= |{\cal A}_E|^2 + |{\cal A}_B|^2,
 \eeq
 which depend on the two-point Wightman functions defined in Eq.~(\ref{eq:wight}) (note that by parity invariance, the mixed correlator $\langle Q^E Q^B\rangle$ vanishes).   For example, the electric term in the case of zero black hole spin is
  \bea
  |{\cal A}_E|^2 = {1\over 4 m_{Pl}^4} {T\over q^4} A_+^E(\omega) \langle a,b|c,d\rangle {\cal A}^E_{ab} {\cal A}^E_{cd},
 \eea
and similarly for the magnetic piece.   The time scale $T=2\pi\delta(\omega\rightarrow 0)$ is an arbitrary IR cutoff associated with time translation invariance which will not appear in physical observables.   We find, from Eqs.~(\ref{eq:AE}),~(\ref{eq:AB}),
\bea
\nn
|{\cal A}_E|^2 &\approx& {T\over 6 m_{Pl}^4} A^E_+(q\cdot v) \left[(v\cdot p)^4 -m^2 (v\cdot p)^2\left(1 - {3\over 2} {(v\cdot q)^2\over q^2}\right)+{1\over 4}m^4 \left(1 + {3} {(v\cdot q)^4\over q^4}\right)\right],\\
\label{eq:a}
\\
\nn
|{\cal A}_B|^2 &\approx& {T\over 8 m_{Pl}^4} A^B_+(q\cdot v) \left[(v\cdot p)^4 - m^2  (v\cdot p)^2\left(1 - 2 {(v\cdot q)^2\over q^2}\right) - m^4 {(v\cdot q)^2\over q^2} \left(1-{(v\cdot q)^2\over q^2}\right)\right],\\
\label{eq:b}
\eea
where we drop terms subleading in powers of the momentum transfer $q$.    The resulting black hole-frame differential cross section for inelastic scattering is then
\bea
\nn
{d^3\sigma \over dq^2 d(q\cdot v)} &\approx& {7 G_N  r_s^5\over 270\pi [(v\cdot p)^2 - m^2]} \left[(v\cdot p)^4  -m^2 (v\cdot p)^2\left(1 - {12\over 7} {(v\cdot q)^2\over q^2}\right) \right.\\
\label{eq:result}
& & \left.+{1\over 7}m^4 \left(1 -3 {(v\cdot q)^2\over q^2}+  6{(v\cdot q)^4\over q^4}\right)\right].
\eea

\begin{figure}
    \centering
    \includegraphics[scale=0.4]{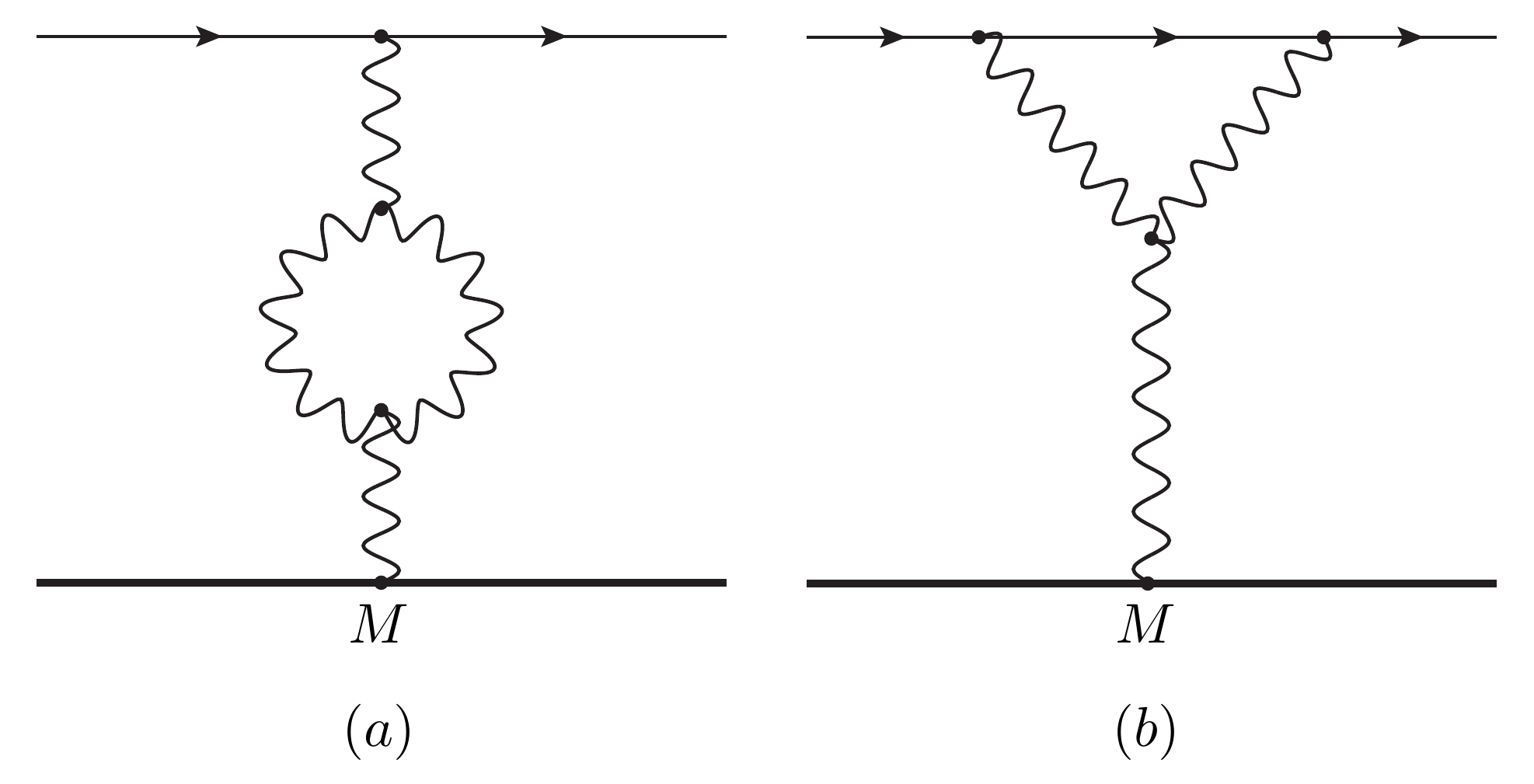}
\caption{Selected one-loop corrections to the elastic scattering process in perturbative quantum gravity.  Diagram (a) contributes at first order in $q^2/m_{Pl}^2\ll 1$, while (b) encodes both classical corrections of order $r_s E_\phi\ll 1$ and quantum effects of order $q^2/m_{Pl}^2$.   The top (arrowed) line corresponds to $\phi$, while the (bottom) solid line corresponds to the black hole, treated as static point source.}
\label{fig:onel}
\end{figure}

It is useful to compare the magnitude of this result with the leading order cross section for Newtonian potential scattering off the black hole.   In the black hole rest frame, this is given by
\beq
{d\sigma_N\over d q^2} = { 4\pi r_s^2\over q^4} {\left[(v\cdot p)^2-{1\over 2} m^2\right]^2\over  (v\cdot p)^2 - m^2}.
\eeq
To compare this to the off-shell Hawking process, we would need to integrate Eqs.~(\ref{eq:a}), Eq.~(\ref{eq:b}) over the region $-\infty<v\cdot q<\infty$.   Although the EFT breaks down when the magnitude of $q\cdot v$ is of order $r_s$, we expect, by unitarity, that the integral over the form factors $A_+^{E,B}(q\cdot v)$ is finite, and dominated by scales near $q\cdot v$.   Thus we may estimate the magnitude of the integrated inelastic (Hawking) differential cross section $d\sigma_H/dq^2$ by taking the result in Eq.~(\ref{eq:result}) and multiplying it by a factor of $q\cdot v\sim r_s^{-1}$.   We then find that, parametrically,
\beq
{d\sigma_H\over d\sigma_N} \sim  {q^2\over m_{Pl}^2},
\eeq
up to factors $(r_s q)^2$ which we cannot determine by purely dimensional arguments and are treated as being of order unity for the purposes of this estimate.   We see that inelastic scattering is a quantum gravity effect, of the same order in  $q^2/m_{Pl}^2$ as the one-loop correction to elastic scattering that arises from graviton vacuum polarization effects of the type first computed in~\cite{tv} and illustrated in Fig.~\ref{fig:onel}.   Our result in Eq.~(\ref{eq:result}) should then be interpreted as a new type of calculable, leading order,  quantum gravity effect in  black hole quantum mechanics. Moreover, the prediction is made within a systematic expansion with calculable
corrections.

\section{Conclusions}
\label{sec:conc}

We have presented what, to our knowledge, is the first computation of quantum gravity effects in scattering processes with black holes appearing as asymptotic states.   Our approach relies on EFT methods presented in~\cite{GnR2,GnR3}.   In this EFT, the leading quantum corrections due to horizon dynamics is represented by the exchange of virtual Hawking radiation.    What is interesting about these effects is that, despite being non-perturbative in nature, are not as suppressed as one might naively have expected.    Instead, they scale in the same way in the $q^2/m_{Pl}^2$ power counting as the more familiar one-loop graviton vacuum polarization~\cite{tv}  corrections to scattering which arise when treating the black hole sources as elementary particles (i.e. quantum fields). 

A natural question to ask is how the inelastic scattering rate calculated here compares to on-shell processes, e.g radiative pressure.
 Given the on-shell nature of the incoming graviton in this case and the fundamental (point-like) nature of the scattered particle, such a process will necessarily be suppressed by further powers of $q^2/m^2_{Pl}$ due to the existence of final state re-radiation.    Similarly, such a process would be suppressed in the case of a particle with non-trivial internal structure, even if no radiation appears in the final state.
 
 In this paper we have only considered a simple inelastic process in which a scalar field scatters gravitationally off a 4D Schwarzschild black hole.  However, our methods should apply more broadly to a larger class of scattering processes as well as to more generic black holes, for instance carrying electric and magnetic charges and/or spin.   Work on such generalizations is underway and will be presented in future publications.

\section{Acknowledgments}
We thank Ted  Jacobson for helpful comments.  This work was partially supported by the US Department of Energy under grants DE-SC00-17660 (WG) and  DE- FG02-04ER41338 and FG02-06ER41449. (IZR).

\appendix

\section{Matching the quadrupole Wightman functions}
\label{app}

In this appendix, we extract the Wightman two-point functions of the quadrupole operators in Eq.~(\ref{eq:EFTS}) by comparing to the the transition probability  $p(n_i\rightarrow n_f)$ obtained in~\cite{bek,wald} to emit $n_f$ identical Hawking bosons, all in the same one-particle state $|\psi\rangle$, from an initial asymptotic state of $n_i$ bosons, also in the state $|\psi\rangle$.

While~\cite{bek,wald} only explicitly considered scalar emission and absorption, their result only relies on canonical quantization of a massless free bosonic field in the background of the black hole, and therefore generalizes to particles with higher spin.   We therefore interpret the result in~\cite{bek,wald},
\beq
\label{eq:bek}
p_{\ell} (n_i\rightarrow n_f) = {(1-x) x^{n_f} (1-|R_{\lambda }|^2)^{n_i+n_f}\over (1 - x |R_{\lambda }|^2)^{n_i+n_f+1}}\sum_{k=0}^{\mbox{min}(n_i,n_f)} {(n_i+n_f-k)!\over k! (n_i-k)! (n_f-k)! } \left[{(|R_{\lambda }|^2-x) (1-x |R_{\lambda }|^2) \over x (1-|R_{\lambda }|^2)^2}\right]^k,
\eeq
as the transition probability for particles in a one-particle wavepacket $|\lambda\rangle$ localized around some energy $\omega$, and carrying definite total angular momentum quantum numbers $\lambda=(\ell,m,h),$ with $\ell\geq 2$, $|m|\leq \ell$ and helicity $h=\pm 2$.   Then $|R_{\lambda }(\omega)|$ is the classical reflection coefficient for the wavepacket $|\lambda\rangle$, obtained in~\cite{page2},  $x=\exp[-\beta_H \omega]$ is the Boltzmann factor for the non-rotating black hole,  $\beta_H=T_H^{-1} = \hbar/4\pi r_s$.  In the limit $\beta_H \omega\ll 1$ in which the EFT is valid, the single particle transition probabilities reduce to
\beq
\label{eq:prob}
p(0\rightarrow 1) \approx p(1\rightarrow 0) \approx {|B_\lambda(\omega)|^2\over\beta_H\omega},
\eeq
where, $|B_\lambda(\omega)|^2=1-|R_{\lambda}(\omega)|^2$  is the classical absorption probability in the state with angular quantum numbers $\lambda$.    

In the EFT, the transition probabilities are 
\beq
p(n_i\rightarrow n_f) = \sum_X |{\cal A}(n_i+M\rightarrow n_f + X)|^2,
\eeq
where, using Eq.~(\ref{eq:EFTS}), the absorption amplitude to leading order in perturbation theory is given
\beq
\label{eq:a10}
i{\cal A}(1+M\rightarrow 0 +X) \approx -i\int dt \langle X|Q^E_{ij}(t)|M\rangle \langle 0|E_{ij}(t,0)|\lambda\rangle + \mbox{magnetic}
\eeq
in the rest frame of the black hole, assumed to be non-rotating.    

To evaluate the matrix element $\langle 0|E_{ij}(t,0)|\lambda\rangle$, we expand in helicity partial waves~\cite{rqm} $|k,\ell,m,h\rangle$ of definite energy $k$, which we normalize as
\beq
\langle k,\ell,m,h|k',\ell',m',h'\rangle = 2\pi\delta(k-k') \delta_{\ell\ell'}\delta_{mm'}\delta_{hh'}.
\eeq
Using the relation between these states and the four-momentum eigenstates $|p,h=\pm 2\rangle$, it is straightforward to show (for instance by working in unitary gauge where only the transverse traceless graviton contributes to on-shell matrix elements):
\beq
\label{eq:ME}
\langle 0|E_{ij}(0)|k,\ell,m,h\rangle = \sqrt{k^5\over 8\pi (2\ell+1) m_{Pl}^2} \delta_{\ell,2} \langle i,j|\ell=2,m\rangle,
\eeq
and $\langle 0|B_{ij}(0)|k,\ell,m,h=\pm 2\rangle=\pm i\langle 0|E_{ij}(0)|k,\ell,m,h\rangle$.    In Eq.~(\ref{eq:ME}), the symbol $ \langle i,j|\ell=2,m\rangle$ denotes the change of basis matrix between Cartesian and spherical rank $\ell=2$ traceless symmetric tensors, with the normalization of the Cartesian states set by $\langle i,j|r,s\rangle={1\over 2} \left[\delta_{ir} \delta_{js} + \delta_{is}\delta_{jr} - {2\over 3} \delta_{ij}\delta_{rs}\right]$.   It follows that for the unit normalized wavepacket $|\lambda\rangle$,
\beq
|\lambda\rangle = \int_0^\infty {dk\over 2\pi} \psi_\lambda(k) |k,\ell,m,h\rangle,
\eeq
with $\int_0^\infty  {dk\over 2\pi} |\psi_\lambda(k)|^2 =1$, the matrix elements are
\beq
\langle 0|E_{ij}(t,0)|\lambda\rangle = {\pi\over \sqrt{5} m_{Pl}} \delta_{\ell,2} \langle i,j|\ell=2,m\rangle \int_0^\infty {k^{5/2} dk\over (2\pi)^{5/2}} e^{-i k t} \psi_\lambda(k),
\eeq
and $\langle 0|B_{ij}(t,0)|\lambda\rangle =\pm i \langle 0|E_{ij}(t,0)|\lambda\rangle$ for $h=\pm 2$ respectively.

Squaring the amplitude in Eq.~(\ref{eq:a10}), we find that after summing over the final states with $\sum_X |X\rangle\langle X|=1$, the single particle absorption probability depends on the two-point functions defined in Eq.~(\ref{eq:wight}).    Given that $|\psi_\lambda(k)|^2$ is sharply localized around $k=\omega$, 
\bea
\nn
p(1\rightarrow 0) &=&  {\pi^2\over 5 m_{Pl}^2}\sum_{r,s,i,j}\int_0^\infty {k^5 dk\over (2\pi)^4}  \left|\psi_\lambda(k)\right|^2 \langle\ell=2,m|r,s\rangle A_+^E(k) \langle r,s|i,j\rangle \langle i,j|\ell=2,m\rangle+\mbox{mag.}\\
\nn\\
&\approx& {4\over 5} G_N \omega^5  \left(A_+^E(\omega) + A_+^B(\omega)\right).
\eea
Similarly, the emission probability is 
\beq
p(0\rightarrow 1)\approx {4\over 5} G_N \omega^5 \left(A_+^E(-\omega) + A_+^B(-\omega)\right).
\eeq
Under the assumption that the magnetic and electric correlators are equal~\cite{GnR2}, comparison to the full theory Eq.~(\ref{eq:prob}), with the $\ell=2$ graybody factor given by~\cite{page2} $|B_\lambda(\omega)|^2\approx {4\over 225} (r_s\omega)^6$ at low energies, then yields the result in Eq.~(\ref{eq:aeab}).      By comparing the EFT to the full theory transition probabilities with more than one particle in the final or initial state, it is possible also to extract the higher-point correlators of the worldline quadrupole operators.   As in~\cite{GnR3}, one would find that the higher-point functions are Gaussian and composed of products of  Schwinger-Keldysh  two-point functions.

\end{document}